\documentclass
[pre,aps,a4paper,twocolumn,twoside,showpacs,showkeys,preprintnumbers,amsfonts,amsmath,amssymb,superscriptaddress,floatfix]{revtex4-1}

\usepackage{graphicx}
\usepackage{dcolumn}
\usepackage{natbib}
\usepackage{bm}
\usepackage{graphics,graphicx}
\usepackage{amsfonts,amsmath,amssymb}
\usepackage{subfigure}
\usepackage{epstopdf}
\usepackage{verbatim}
\usepackage{times}

\usepackage{hyperref}

\begin{document}

\title{Flow-induced shear instabilities of cohesive granulates}

\author{Ilenia Battiato}\email{ibattia@clemson.edu}  
\affiliation{Clemson University, Department of Mechanical Engineering, Clemson, SC 29634, USA}
\affiliation{Max Planck Institute for Dynamics and Self-Organization (MPIDS), 37077 G\"{o}ttingen, Germany}

\author{J\"{u}rgen Vollmer}
\affiliation{Max Planck Institute for Dynamics and Self-Organization (MPIDS), 37077 G\"{o}ttingen, Germany}
\affiliation{Fakult\"at f\"ur Physik, Universit\"at G\"ottingen, 37077 G\"ottingen, Germany}

\date{20 March 2012}

\newcommand\dynpercm{\nobreak\mbox{$\;$dyn\,cm$^{-1}$}}
\newcommand\cmpermin{\nobreak\mbox{$\;$cm\,min$^{-1}$}}

\providecommand\bnabla{\boldsymbol{\nabla}}
\providecommand\bcdot{\boldsymbol{\cdot}}
\newcommand\biS{\boldsymbol{S}}
\newcommand\etb{\boldsymbol{\eta}}

\newcommand\Real{\mbox{Re}} 
\newcommand\Imag{\mbox{Im}} 
\newcommand\Rey{\mbox{\textit{Re}}}  
\newcommand\Pran{\mbox{\textit{Pr}}} 
\newcommand\Pen{\mbox{\textit{Pe}}}  
\newcommand\Ai{\mbox{Ai}}            
\newcommand\Bi{\mbox{Bi}}            

\newcommand\ssC{\mathsf{C}}    
\newcommand\sfsP{\mathsfi{P}}  
\newcommand\slsQ{\mathsfbi{Q}} 

\newcommand\hatp{\skew3\hat{p}}      
\newcommand\hatR{\skew3\hat{R}}      
\newcommand\hatRR{\skew3\hat{\hatR}} 
\newcommand\doubletildesigma{\skew2\tilde{\skew2\tilde{\Sigma}}}

\newsavebox{\astrutbox}
\sbox{\astrutbox}{\rule[-5pt]{0pt}{20pt}}
\newcommand{\astrut}{\usebox{\astrutbox}}

\newcommand\GaPQ{\ensuremath{G_a(P,Q)}}
\newcommand\GsPQ{\ensuremath{G_s(P,Q)}}
\newcommand\p{\ensuremath{\partial}}
\newcommand\tti{\ensuremath{\rightarrow\infty}}
\newcommand\kgd{\ensuremath{k\gamma d}}
\newcommand\shalf{\ensuremath{{\scriptstyle\frac{1}{2}}}}
\newcommand\sh{\ensuremath{^{\shalf}}}
\newcommand\smh{\ensuremath{^{-\shalf}}}
\newcommand\squart{\ensuremath{{\textstyle\frac{1}{4}}}}
\newcommand\thalf{\ensuremath{{\textstyle\frac{1}{2}}}}
\newcommand\Gat{\ensuremath{\widetilde{G_a}}}
\newcommand\ttz{\ensuremath{\rightarrow 0}}
\newcommand\ndq{\ensuremath{\frac{\mbox{$\partial$}}{\mbox{$\partial$} n_q}}}
\newcommand\sumjm{\ensuremath{\sum_{j=1}^{M}}}
\newcommand\pvi{\ensuremath{\int_0^{\infty}%
  \mskip \ifCUPmtlplainloaded -30mu\else -33mu\fi -\quad}}

\newcommand\etal{\mbox{\textit{et al.}}}
\newcommand\etc{etc.\ }
\newcommand\eg{e.g.\ }

\newcommand{\e}{\mathrm{e}}
\newcommand{\Pe}{\mathrm{Pe}}
\newcommand{\Da}{\mathrm{Da}}
\newcommand{\Ca}{\mathrm{Ca}}
\newcommand{\M}{\mathrm{M}}
\newcommand{\de}{\mathrm{d}}
\newcommand{\ee}{\mathbf e}
\newcommand{\bb}{\mathbf b}

\begin{abstract}
  In this work we use a multi-scale framework to calculate the
  fluidization threshold of three-dimensional cohesive granulates
  under shear forces exerted by a creeping flow. A continuum model of
  flow through porous media provides an analytical expression for the
  average drag force on a single grain. The balance equation for the
  forces and a force propagation model are then used to investigate
  the effects of porosity and packing structure on the stability of
  the pile.  We obtain a closed-form expression for the instability
  threshold of a regular packing of mono-disperse frictionless
  cohesive spherical grains in a planar fracture. Our result
  quantifies the compound effect of structural (packing orientation
  and porosity) and dynamical properties of the system on its
  stability.
\end{abstract}



\maketitle

\section{Introduction}

Granulates are a large collection of macroscopic solid grains. Dry and
wet granulates are vital in a large variety of industries, ranging
from pharmaceutical to mining \citep[][pp. 4-10]{duran}, from
construction \cite{Watano2003} to agricultural
\citep[][pp. 10]{duran}. They also play an important role in many
geological processes, such as land and mudslides \citep{dikau}, debris
flows \citep{iverson-2000-acute}, erosion, particles’ resuspension by
wind in humid regions \citep{Nicholson1988}, and dune formation
\citep{bagnold} that shape planets' morphology including, but not
limited to, Earth \citep{hansen-2011-mars}.

Most theoretical and experimental studies focus on dry granulates and
their collective behaviour including pattern formation
\citep[e.g.,][]{umbanhowar-1996-localized,hansen-2011-mars}, angle of
stability/repose
\citep[e.g.,][]{jaeger-1989-relaxation,alonso-1996-shape}, avalanches
dynamics \citep[e.g.,][]{jop-2006-initiation,jop-2007-initiation} and
granular flows
\citep[e.g.,][]{rajchenbach-2003-dense,gdr-2004-dense}. However, as
every child knows, adding even a small quantity of liquid to a
sandpile dramatically changes its properties.

Cohesive interactions due to capillarity exist in three-phase systems
such as partially-wet granulates where solid grains, wetting and
non-wetting fluids (e.g. water and air) coexist.  The existence of a
\textit{cohesive force} between grains leads to fundamentally
different dynamics in wet granulates compared to their dry
(i.e. non-cohesive) counterpart. Such differences include stability of
granular piles and the location in a granular bed where incipient
motion, either due to gravity
\citep{Novak-2005-maximum,iverson-2000-acute,rahbari-2009-response} or
shearing
\citep[e.g.,][]{Charru:2002kx,Mouilleron:2009vn,Ouriemi:2009ys,Rahbari:2010fk},
is first observed.

The collective behaviour of cohesive grains has only recently begun to
be explored.  A number of studies have focused on the dynamics of wet
granular avalanches \citep[e.g.,][]{tegzes-2003-development} and on
the effect of humidity
\citep{Fraysse-1999-humidity,Fray-2001-humidity,restagno-2002-aging}
and capillary forces
\citep{hornbaker-1997-sandcastle,restagno-2004-cohesive,Novak-2005-maximum}
on the static properties of granulates in both engineering
applications \citep{Fray-2001-humidity} and natural systems, e.g. soil
\citep{iverson-2000-landslide}.  A number of different models have
been proposed to study the geometric stability of wet piles, including
Mohr-Coulomb continuum \citep{mason-1999-critical}, liquid-bridge
\citep{Novak-2005-maximum} and response function
\citep{Mourzakel-1998-isostatic,Moukarzel-2004-book,rahbari-2009-response}
models.

While such models are invaluable in shedding light into the properties
of \textit{cohesive} granulates, natural systems often include a
number of additional forcing factors that might significantly affect
the stability of granulates in the environment. In geological systems,
instability is triggered by a combination of body forces
(e.g. gravity) and hydrodynamic shearing due to the creeping motion of
a fluid through the granulate.  This is especially true in processes
such as cliff instability and landslides, sediment transport in
submerged environments (e.g.  seafloor transport), and fluidization of
fines in fractures during pumping operations or oil recovery, just to
mention a few. Even though the understanding of how flow-induced shear
forces affect the stability of granular matter is of utmost importance
to better quantify the processes that trigger a fluidization event,
incorporating such effects is a challenging task since it requires the
solution of Navier-Stokes equations in highly complex geometries.

\begin{figure}
 \centerline{\includegraphics[width=0.5\textwidth]{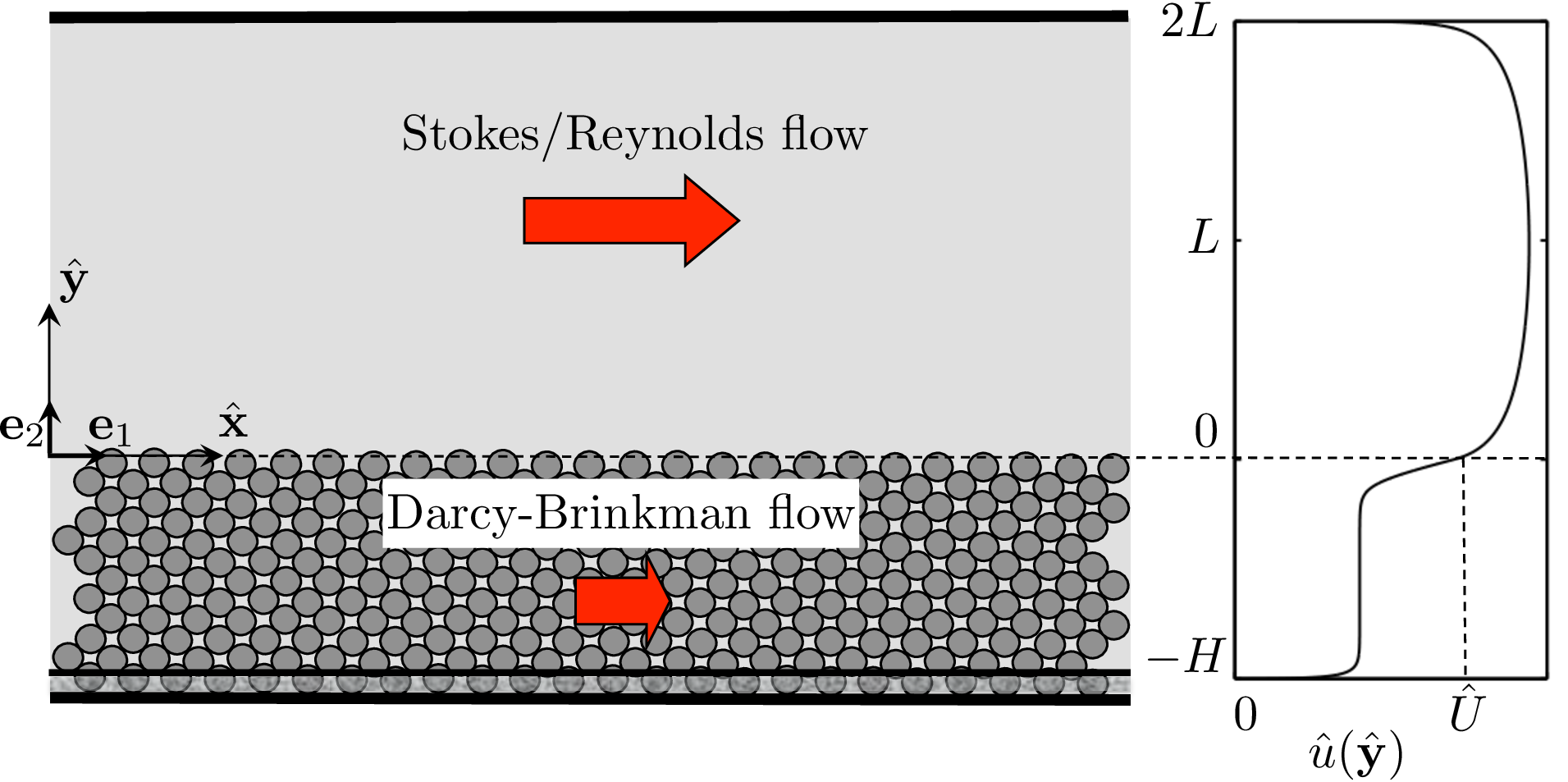}}
  \caption{Schematic of the domain on the left, and shape of the average velocity profile through the channel, on the right. The grains are assumed to be \textit{cohesive}.}\label{fig:domain}
\end{figure}

In the present work we address this problem in a multi-scale framework
and quantify the effect of dynamic shearing forces due to the creeping
flow of a fluid (e.g. air) on the onset of instability
(i.e. fluidization) of a cohesive granular pile
(Fig.~\ref{fig:domain}). Explicit analytical solutions are obtained
for a model setting where the granulate is constituted of
mono-disperse, frictionless, \textit{cohesive} grains arranged in a
regular packing. The cohesive interactions are due to the presence of
capillary bridges formed by a wetting fluid, e.g. water, at the
contact points between grains. Complications associated with random
packing and friction between particles need not be taken into account
to obtain reasonable results, as shown in \citep{Novak-2005-maximum}.

In section~\ref{sec:formulation}, we treat the cohesive granulate as a
porous medium, and introduce a continuum-scale Darcy-Brinkmann model
for the flow and the average drag force exerted by the fluid on the
grains.  Section~\ref{sec:pore-scale-model} discusses a pore-scale
network model for the force propagation through the pile. Location of
failure and maximum load are derived. In
section~\ref{sec:fluidization} the stability criterion is formulated
in terms of the capillary number, that represents the relative
strength between destabilising flow-induced shear and stabilising
capillary (cohesive) forces, and the packing orientation relative to
the average flow direction.  The main results and conclusions are
summarised in section~\ref{sec:conclusions}. For simplicity, gravity
is here neglected. Generalisation to include gravity effects is
straightforward.

\section{Continuum-scale Model of Flow and Drag in a Brinkman medium}
\label{sec:formulation}

We consider a fully developed incompressible fluid flow, e.g. air,
between two infinite parallel plates separated by the distance of
$H+2L$.  The bottom part of the flow domain, $-H<\hat y <0$, is
occupied by a packing of cohesive (e.g. water-wet) mono-disperse rigid
frictionless spheres of radius $R$.  The (air) flow is driven by an
externally imposed (mean) constant pressure gradient $\mathrm d_{\hat
  x}\hat p <0$. Therefore, each spherical grain is subject to a drag
force due to aerodynamic stresses and attractive capillary bridge
forces.

Since we are concerned with the fluidization threshold of the sphere
packing, initially at rest, we treat the sphere-packed region as a
porous medium with porosity $\phi$ and constant permeability $K$.
This allows us to decouple an analysis of the flow from that of the
granulate dynamics.  We allow the flow over the granulate to span both
laminar and turbulent regimes. Channel flow through, $\hat
y\in(-H,0)$, and over, $\hat y\in(0,2L)$, a permeable layer can be
described by coupling Brinkman with Navier-Stokes or Reynolds equation
for the horizontal component $\hat u(\hat y)$ of the average velocity
$\hat{\mathbf u}(\hat u, \hat v)$
\citep{battiato-2010-elastic,battiato-2011-universal}
\begin{subequations}
\begin{eqnarray}
& \mu_e \de_{\hat y\hat y} \hat{u}
 - \mu K^{-1}\hat{u}-\de_{\hat x}
 \hat p=0\quad &  \quad \hat y\in(-H,0), \label{eq:brinkman_x}\\
& \mu \de_{\hat y\hat y} \hat{u} - \rho \gamma
\de_{\hat y} \langle \hat u' \hat v'\rangle-\de_{\hat x}
 \hat p=0 & \quad  \hat y \in(0,2L), \label{eq:channel}
\end{eqnarray}
\end{subequations}
where $\de_{\hat x} \hat p$ is a mean constant pressure gradient,
$\mu$ and $\rho$ are the fluid's dynamic viscosity and density,
respectively, and $\mu_e$ is its ``effective" viscosity that accounts
for the slip at the spheres walls. In the laminar regime ($\gamma=0$),
$\hat u$ is the actual velocity and $\hat v \equiv 0$. In the
turbulent regime ($\gamma=1$), the actual velocity is decomposed into
a mean velocity $\hat{\mathbf u}$ and velocity fluctuations $\hat u'$
and $\hat v'$ about their respective means.  $\langle \hat u' \hat
v'\rangle$ denotes the Reynolds stress. Fully-developed turbulent
channel flow has velocity statistics that depend on $\hat y$ only.

In both laminar and turbulent regimes, the no-slip condition requires
zero velocity at $\hat y=-H$ and $\hat y=2L$, and the continuity of
velocity and shear stress is prescribed at the interface, $\hat y=0$,
between the free and filtration flows \citep{vafai-1990-fluid}:
\begin{eqnarray}\label{bcs:dimensional}
& \hat{u}(-H) =\hat{u}(2L) =0, \quad 
\hat u(0^-)=\hat{u}(0^-) = \hat U, \nonumber \\
& \mu_e  \left.\de_{\hat y} \hat{u} \right|_{0^-}=  \mu
\left.\de_{\hat y} \hat{u}\right|_{0^+}
\end{eqnarray}
where $\hat U$ is an unknown matching velocity at the interface between channel flow and porous medium.

Choosing $(\mu,H,q)$, with $q = - \mu^{-1}H^2 \de_{\hat x} \hat p$ a characteristic Darcy velocity, as the repeating variables, the problem can be formulated in dimensionless form. Then, inside the granular medium, the solution for the dimensionless velocity distribution $u=\hat u/q$ is given by   \citep{battiato-2010-elastic}
\begin{subequations}\label{eq:brink-vel}
\begin{equation}\label{eq:brink-velocity-only}
u(y) = \M^{-1}\lambda^{-2}+C_1 \e^{\lambda y} + C_2 \e^{-\lambda y}, \qquad y\in(-1,0),
 \end{equation}
 where $y=\hat y /H$, $\M=\mu_e/\mu$,  $\delta=L/H$, $\lambda^2 = H^2/(\M K)$, and 
 \begin{eqnarray}\label{brinkman:constants}
&C_{1,2} =\pm\dfrac{1}{\M\lambda^2}\dfrac{(\M\lambda^2U-1)\e^{\pm\lambda}+1}{
 \e^\lambda-\e^{-\lambda}}, \\  
 &U = \dfrac{1}{\beta\M\lambda^2}\left(1- \mbox{sech} \lambda+ \delta\lambda\tanh \lambda\right),\label{U-def}
 \end{eqnarray}
\end{subequations}
with $U=\hat U/q$  the dimensionless interfacial velocity, and  $\beta=1$ or $\beta=1+(\tanh \lambda) /( 2\delta \M \lambda)$ for turbulent or laminar regime in the channel, respectively. In the following, we set $\M=1$ since the fluid does not experience any slip on the grains' walls.

The total drag on a sphere in an unbounded Brinkman medium is given by \citep{howells-1974-drag,hinch-1977-averaged,1985-kim-modelling}
\begin{align}\label{eq:drag}
\hat{\mathbf F} =6 \pi \mu  R  g(\phi)\hat{\mathbf V},
\end{align}
where $\hat{\mathbf V}$ is a uniform velocity at infinity,
$g(\phi)=1+\frac{3}{\sqrt{2}}(1-\phi)^{1/2}+\frac{135}{64}(1-\phi)\ln(1-\phi)+16.456(1-\phi)+o(1-\phi)$
\citep[p.508, eq. (19.119)]{Torquato} and $\phi$ is porosity.
Permeability, obtained by self-consistent arguments, is given by
$K=k_s g^{-1}(\phi)$ \citep{howells-1974-drag,hinch-1977-averaged},
where $k_s=\frac{2}{9}R^2(1-\phi)^{-1}$ is the well-known Stokes
result, for low-porosity packing of spheres.
Since non-uniform velocity effects in
Eq. \eqref{eq:brink-velocity-only} are confined to a small region
close to the upper and lower boundaries of the granulate, we employ a
vertically averaged velocity $\bar{u}(y)$ to calculate a first-order
approximation of the drag.  Therefore, combining
Eqs. \eqref{eq:brink-velocity-only} and \eqref{eq:drag}, the
dimensionless drag force $\mathbf F(y):=(\mu q H)^{-1}\hat{\mathbf
  F}=[F(y), 0, 0]=F(y) \mathbf e_1$ exerted by the fluid on a sphere
centered at $y$, is given by
\begin{align}\label{eq:drag-dimensionless}
F(y)=3\pi\epsilon g(\phi)\bar{u}(y)
\end{align}
where $\epsilon=2R/H$ is the dimensionless grain diameter,
$\bar{u}(y)=\frac{1}{2\bar h}\int_{y-\bar{h}}^{y+\bar{h}} u(y')\de y'$
is an average velocity across a layer of thickness $2\bar h$, and
$\mathbf{e}_1$ is the unit vector in the $x$-direction.

In the following section we specialize the analysis to a regular
(cubic) packing of spheres. This will allow us to determine the
network of forces, and consequently the maximum load, developed inside
the pile.

\begin{figure}
  \centerline{\includegraphics[width=0.5\textwidth]{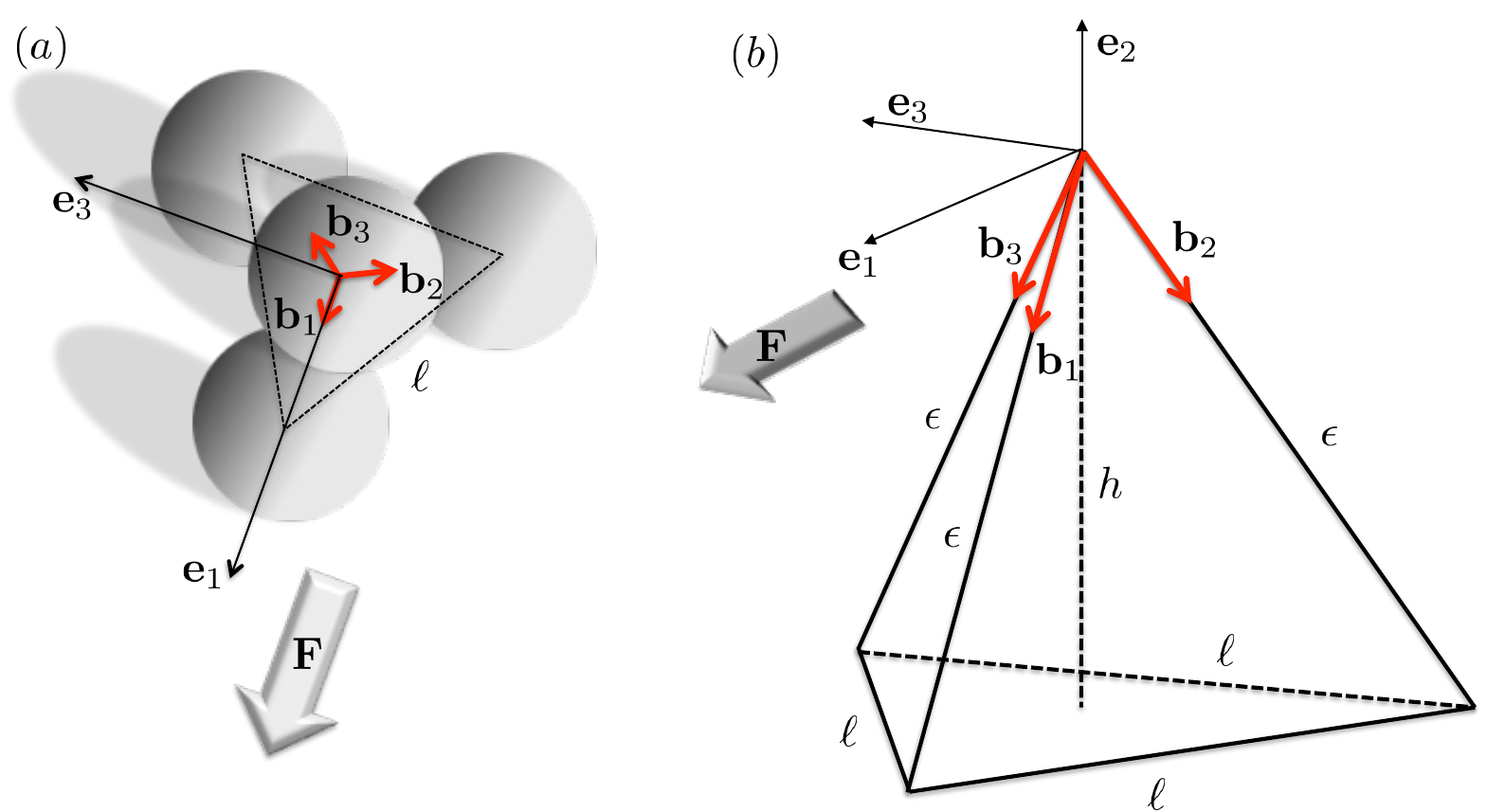}}
  \caption{(a) Top view of the structure of a two-layer 3-dimensional
    pile of spheres of dimensionless diameter $\epsilon$. (b) Sketch
    of the tetrahedron obtained by connecting the centers of the
    4-sphere structure (left).  $\ell$ and $h$ represent the
    intralayer and iterlayer distances between sphere centers
    belonging to either the same or adjacent layers,
    respectively. $\mathbf{e}_\alpha$, $\alpha=\{1,2,3\}$, are unit
    vectors of the canonical orthonormal basis of the Euclidean
    space. $\mathbf{b}_\alpha$, $\alpha=\{1,2,3\}$, are unit vectors
    along the lattice directions connecting the center of the
    supported sphere with the centers of the supporting
    spheres.}\label{fig:thetrahedron}
\end{figure}
\section{Force Network model and maximum load}
\label{sec:pore-scale-model}

\subsection{Geometry and packing}

Let the mono-disperse cohesive grains be arranged in an isostatic
packing, obtained by expanding a face-centered cubic packing so to
eliminate interlayer contacts, with the (111)-face of the crystal
parallel to the bottom wall of the channel. Such expanded packing
configuration will be referred to as cubic expanded packing
(CEP). From the first two layers of spheres (AB), a CEP arrangement
can be obtained if every third layer is the same
\citep{ashcroft-book}. Figures~\ref{fig:thetrahedron}(a) and (b) show
the top and side views of the structure of the first two layers of
spheres. While we focus on such a specific grains' arrangement, we
stress that the analysis can be easily generalised to other regular
packing structures. Let $\hat \ell$, with $2R<\hat\ell<2\sqrt{3}R$, be
the pitch in the $x-z$ (horizontal) plane, i.e. the distance between
the centers of spheres belonging to the same layer, and $\hat h$ the
pitch in the $y-z$ (vertical) plane, i.e. the distance between two
adjacent layers (Fig.~\ref{fig:thetrahedron}). The dimensionless
interlayer and intralayer distances $h=\hat hH^{-1}$ and $\ell=\hat
\ell H^{-1}$, respectively, are related as follows
$h=\epsilon[1-\frac{1}{3}( \ell/\epsilon)^2]^{1/2}$. Porosity, $\phi$,
amounts to
$\phi=1-\pi[3(\ell/\epsilon)^2\sqrt{3-(\ell/\epsilon)^2}]^{-1}$, with
$\epsilon<\ell<\sqrt{3}\epsilon$. When $\ell=\epsilon$ the close
packing is recovered and $\phi \rightarrow \phi_c\approx 0.26$,
corresponding to the face-centered cubic packing fraction of spheres,
$s_c=\pi/3\sqrt{2}\approx 0.74$. If $\ell\rightarrow\sqrt{3}
\epsilon$, $h\rightarrow 0$ and the four spheres lie on the same
level.

\begin{figure}
 \centerline{\includegraphics[width=0.5\textwidth]{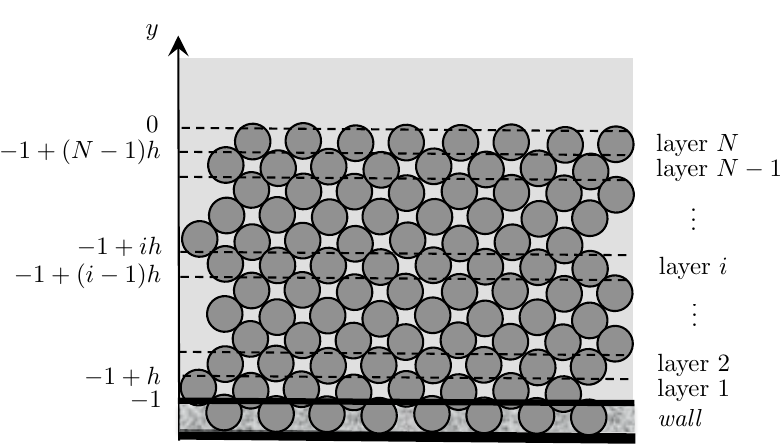}}
 \caption{Schematics of the layer enumeration. The dashed horizontal
   lines represent the location of the contacts between two adjacent
   layers of grains.}\label{fig:pore-continuum}
\end{figure}

Let $N+1$ be the total number of layers in the pile. The bottom layer
of grains is immobile and is called a \textit{wall}. Therefore only
$N$ layers are mobile. Let $n=\{1,\cdots,N\}$ denote the (mobile)
layer number.  The layer sitting immediately on the \textit{wall} has
$n=1$. The layer enumeration continues moving up to the top-most layer
in the pile where $n=N$ (see Fig.~\ref{fig:pore-continuum}).  The
number of mobile layers, $N$, and the height of the pile, $H$, are
related through $H=N\hat h$ or in terms of dimensionless quantities
$N=1/h$.

In the following section a pore-scale force network model is used to
determine the location and the modulus of the maximum load in the
pile.

\subsection{Propagation of forces}

Let $\mathbf F_k$ be the sum of the external forces acting on grain
$k$ (e.g. drag), and $\mathbf g_{k l}=g_{kl}\bb_{kl}$ the force
exerted from grain $k$ to grain $l$, where $\bb_{kl}$ is a unit vector
pointing from grain $k$ to grain $l$ and $g_{kl}$ is the magnitude of
the force; $g_{kl}$ is positive for compressive forces and negative
otherwise.  Whenever there is a stretched capillary bridge between two
grains $k$ and $l$, the force exerted by grain $k$ on $l$ is
attractive and equal to a constant value $\mathbf f_{kl}=f_b\bb_{kl}$
where $f_b>0$. This simplifying assumption that the capillary force is
a constant, irrespective of grain separation distance, has
demonstrated to provide good description of collective behaviour of
wet
granulates~\citep{rahbari-2009-response,fingerle-2008-phase,ulrich-2009-cooling}.
The force distribution can be uniquely determined by solving the
following system for the unknowns $g_{kl}$
\begin{eqnarray}\label{eq:general-balance0}
\sum_l \mathbf g_{kl}=\mathbf F_k, \quad \forall k,l=1,..., M,
\end{eqnarray}
with $M$ the total number of grains. 
There is a unique solution for the force distribution~\citep{Mourzakel-1998-isostatic,Moukarzel-2004-book,rahbari-2009-response} in $d$ dimensions, if the packing is isostatic, i.e., if the average number  $\nu$ of neighbours per grain equals $2d$.
Note, however, that this solution does not necessarily comply with the constraint that contacts break when tensile forces exceed the capillary bridge force. Upon variation of parameters a regular packing yields when this additional requirement is first violated, i.e., the packing is stable as long as $ g_{kl}+f_b>0$, $\forall k,l$. 
Specifically, in a regular isostatic 3-dimensional CEP packing of grains bounded by a solid
wall at the bottom and a free surface at the top  each sphere is
supported by $d=3$ contacts from grains further down in the pile. For any couple $(k,l)$ of grains  in contact, the unit vector $\mathbf b_{kl}$ connecting their centers is aligned to one of the (three) CEP-lattice directions.
The external force (e.g. drag) exerted on each grain up in the pile propagates unchanged down
through the pile to the first mobile layer ($n=1$) along such directions as discussed in
\cite{rahbari-2009-response}. Therefore, the maximum load is
experienced from grains at the bottom of the pile, i.e. $n=1$, and
the stability threshold is determined by the respective bonds carrying the highest load. Next, we calculate the maximum destabilising force exerted on the lower-most grains in the pile.

\subsection{Maximum load}

From the continuum model solution, the average drag force $\mathbf
F(n)=[F(n), 0, 0]=F(n)\mathbf e_1$ exerted by the fluid on a grain
belonging to layer $n$ can be obtained by setting the averaging
interval in Eq. \eqref{eq:drag-dimensionless} equal to the layer
thickness, i.e. $2\bar h=h$,
\begin{align}\label{eq:force-layer}
F(n)=\dfrac{3\pi \epsilon}{h}g(\phi)\int_{-1+(n-1)h}^{-1+nh}u(y) \de y, \,\, n=1,\cdots,N.
\end{align}
where $u(y)$ is given by  Eq. \eqref{eq:brink-vel}. Therefore, 
\begin{align}\label{eq:force-layer-calculated}
 F(n)=\dfrac{3\pi \epsilon}{\lambda h}g(\phi)\left[\dfrac{h}{\M\lambda}+C_1\e^{\lambda (-1+nh)}\left(1-\e^{-\lambda h}\right)+\right.\nonumber \\
\left. -C_2\e^{\lambda (1-nh)}\left(1-\e^{\lambda h}\right) \right],
\end{align}
for $n=1,\cdots,N,$ which gives the drag distribution due to aerodynamic shear exerted by the creeping fluid on the wet granular bed. The maximum load $\bar{\mathbf F}$, exerted on the grains of the first layer ($n=1$), is
\begin{align}\label{eq:total_F_CEP}
\bar{\mathbf F}=\sum_{n=1}^N \mathbf F(n).
\end{align}
Combining Eqs. \eqref{eq:total_F_CEP} and \eqref{eq:force-layer} or  \eqref{eq:force-layer-calculated} we obtain
\begin{align}\label{eq:total_force}
\bar{\mathbf F}=\mathbf e_1\dfrac{3\pi \epsilon}{h}g(\phi) \int_{-1}^{0}u(y) \de y,
\end{align}
which gives
\begin{align}\label{eq:total_force2}
\bar{\mathbf F}=\dfrac{3\pi \epsilon}{h}g(\phi)  U_\mathrm{av} \mathbf e_1,
\end{align}
where
\begin{align}
  U_\mathrm{av}=\dfrac{1}{\M \lambda^3}\left[\lambda+\left(\M\lambda^2 U-2\right)\left(\coth\lambda-\mbox{csch}\lambda\right)\right],
\end{align}
is the average velocity across the granulate, and  $U$ is given by Eq.~\eqref{U-def} for
laminar and turbulent regimes above the granulate.

\section{Fluidization threshold}
\label{sec:fluidization}

While the maximum destabilizing average force $\bar{\mathbf F}$ on
each grain in the bottom layer due to Stokes flow in the pile is
parallel to the channel boundary, the stabilizing capillary forces act
along the lattice directions $\{\bb_1,\bb_2,\bb_3\}$, see
Fig.~\ref{fig:thetrahedron}(b). Therefore, if the components of the
total force $\bar{\mathbf{F}}$ along such directions are less than the
capillary forces (assumed constant), the pile is stable: the stability
criterion can be formulated by decomposing the force $\bar{\mathbf F}$
onto the non-orthogonal basis $\{\bb_1,\bb_2,\bb_3\}$ uniquely
identified by the structure of the
packing.

In the following, we proceed with the non-orthogonal projection of the
maximum load which allows us to analytically calculate the instability
threshold while incorporating the impact of the lattice orientation
relative to the average flow direction. We stress that such approach
is readily generalizable to other packing structures and to
incorporate any type of de/stabilizing forces (e.g. gravity,
friction).

\subsection{Non-orthogonal projection}

Let $\mathcal{E}=\{\ee_1, \ee_2, \ee_3\}$ be the canonical orthonormal
basis of the Euclidean space $\mathbb{R}^3$ and $\mathcal{B}=\{\bb_1,
\bb_2, \bb_3 \}$ a generally non-orthogonal basis with $\bb_\alpha$
unit vectors, and $\alpha=\{1,2,3\}$. Let $F_\alpha$ be the components
of the maximum force $\bar{\mathbf F}$ in the canonical basis, and
$F'_\alpha$ its components in the basis $\mathcal{B}$, i.e.
\begin{eqnarray}
\bar{\mathbf F} = \sum_{\alpha=1}^3 F_\alpha \ee_\alpha=\sum_{\alpha=1}^3 F'_\alpha \bb_\alpha.
\end{eqnarray}
The components of $\bar{\mathbf F}$ in the two basis are related through a linear transformation $\mathbf A$,
\begin{eqnarray}\label{eg:change-basis}
(F_1,F_2,  F_3)=\mathbf A (F'_1, F'_2,  F'_3)
\end{eqnarray}
with $(F_1,F_2, F_3)$ and $(F'_1, F'_2, F'_3)$ column vectors, and
$\mathbf A$ the matrix of direction cosines whose components are
defined as $A_{\alpha\beta}=\cos(\bb_\beta,\ee_\alpha)=\bb_\beta\cdot
\ee_\alpha$.

In the $\mathcal{B}$-coordinate system and for any sphere belonging to
the first mobile layer (i.e. $n=1$), Eq. \eqref{eq:general-balance0}
simplifies to
\begin{eqnarray}\label{eq:components}
F'_\alpha=f_b \quad \alpha=\{1,2,3\},
\end{eqnarray}
and the stability criterion is
\begin{eqnarray}\label{eq:stability-generalized}
F'_\alpha\leq f_b, \quad \alpha=\{1,2,3\}.
\end{eqnarray}
Combining Eqs. \eqref{eg:change-basis} and \eqref{eq:stability-generalized} yields to
\begin{eqnarray}\label{stability1-inequality}
B_{\alpha\beta} F_\beta\leq f_b, \quad \alpha,\beta=\{1,2,3\},
\end{eqnarray}
where $B_{\alpha\beta}$ are the components of $\mathbf B:=\mathbf
A^{-1}$. If the components of the maximum load $F_\beta$ satisfy the
system of (three) equations \eqref{stability1-inequality}, then the
pile is stable. Alternatively, Eq.~\eqref{stability1-inequality} can
be solved for the unknowns $F_\beta$, which provide the maximum
magnitude of the components of the load that the capillary forces in
the bottom layer can sustain.
\begin{figure}
  \centerline{\includegraphics[width=0.5\textwidth]{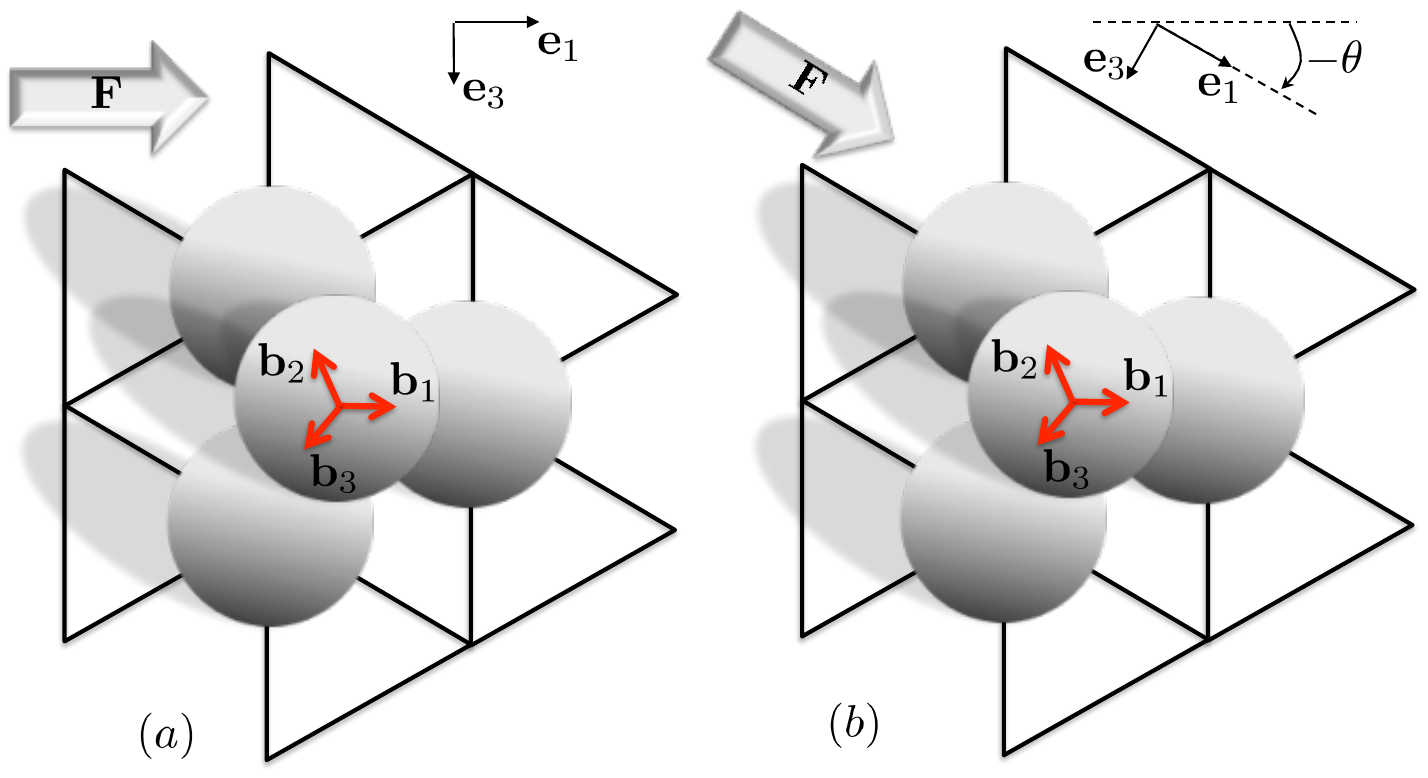}}
  \caption{Top view of the bottom two layers of a regular isostatic
    packing of mono-disperse spheres before (a) and after (b) a
    counterclockwise rotation $\theta$ of the
    packing.}\label{fig:Spheres-pattern}
\end{figure}

The effect of the packing orientation on the pile stability can be
readily incorporated. Without loss of generality, let us consider a
counterclockwise rotation of the pile (and therefore of the basis
$\mathcal{B}$) around the wall-normal, i.e.  $y$-axis (see
Figures~\ref{fig:Spheres-pattern} (a) and (b)). Such solid-body
rotation is fully described by the rotation matrix
$\mathbf{R}_{y}(\theta)$ defined as

\begin{eqnarray}
\mathbf{R}_{y}(\theta) =\left[
\begin{array}{ccc}
\cos\theta & 0 & \sin\theta\\
0 & 1 & 0\\
-\sin\theta & 0 & \cos\theta\\
\end{array}
\right]
\end{eqnarray}
where $\theta$ is the rotation angle. The matrix of direction cosines for the rotated system, $\mathbf A_\theta$, is
\begin{eqnarray}
\mathbf A_\theta=\mathbf{R}_{y}(\theta) \mathbf A.
\end{eqnarray}
The stability criterion, given by Eq. \eqref{eq:stability-generalized}, now implies
\begin{eqnarray}\label{eq:stability_criterion}
B_{\theta,\alpha \beta} F_\beta \leq f_b
\end{eqnarray}
where $B_{\theta,\alpha \beta}$ are the components of $\mathbf
B_\theta =(\mathbf A_\theta)^{-1}$, which gives a generalized
stability criterion for an arbitrary orientation of the packing
structure relative to the average direction of the aerodynamic drag.

\subsection{Stability Diagram}
\label{sec:Fluidization Threshold}

Let us consider the Cartesian coordinate system as depicted in
Fig.~\ref{fig:domain}, where the axis $\ee_1$ and $\ee_2$ of the
Cartesian basis $\mathcal{E}$ are parallel and orthogonal to the
fracture boundary.  We take the projection of $\bb_1$ onto the
$xz$-plane parallel to $\ee_1$ as a reference configuration for the
packing orientation (Fig.~\ref{fig:Spheres-pattern}(a)).
Therefore, the components of the basis vectors $\{\bb_1,\bb_2,\bb_3\}$ in the canonical basis $\mathcal{E}$ and the matrix of direction cosines $\mathbf A$ are
\begin{eqnarray}
&\bb_{1}=\dfrac{1}{\epsilon}\left[
\begin{array}{c}
\frac{\sqrt{3}}{3}\ell \\ \\
-h \\ \\
0 \\
\end{array}
\right], \, \, 
\bb_{2}=\dfrac{1}{\epsilon}\left[
\begin{array}{c}
-\frac{\sqrt{3}}{6}\ell \\ \\
-h \\ \\
-\frac{1}{2} \ell\\
\end{array}
\right], \,\,
\bb_{3}=\dfrac{1}{\epsilon}\left[
\begin{array}{c}
-\frac{\sqrt{3}}{6}\ell \\ \\
-h \\ \\
\frac{1}{2}\ell \\ 
\end{array}
\right], \nonumber \\ \nonumber \\
 &\mathbf A =\dfrac{1}{\epsilon}\left[
\begin{array}{ccc}
\frac{\sqrt{3}}{3}\ell  & -\frac{\sqrt{3}}{6}\ell  &  -\frac{\sqrt{3}}{6}\ell\\ \\
-h & -h & -h\\  \\
0 & -\frac{1}{2}\ell & \frac{1}{2}\ell\\
\end{array}
\right].
\end{eqnarray}
The matrix of direction cosines $\mathbf A_\theta$ after a
counterclockwise rotation of angle $\theta$ about $\ee_2$-axis
(Fig.~\ref{fig:Spheres-pattern} (b)) has the following components
\begin{eqnarray}
\left[\mathbf A_\theta\right]_{ij}=\dfrac{1}{\epsilon}A_{\theta,ij}
\label{eq:Atheta}
\end{eqnarray}
where 
\begin{align}
& A_{\theta,11}=\frac{\sqrt{3}}{3}\ell \cos\theta,  \nonumber \\
 & A_{\theta,12}= -\frac{\sqrt{3}}{6}\ell \cos\theta -\frac{\ell}{2}\sin\theta, \nonumber \\
& A_{\theta,13}= -\frac{\sqrt{3}}{6}\ell \cos\theta +\frac{\ell}{2}\sin\theta, \nonumber \\
& A_{\theta,21}=  A_{\theta,22}= A_{\theta,23}=-h, \label{eq:Atheta-components} \\
& A_{\theta,31}=-\dfrac{\sqrt{3}}{3}\ell \sin\theta, \nonumber \\ 
&   A_{\theta,32}=\dfrac{\sqrt{3}}{6}\ell \sin\theta -\dfrac{\ell}{2}\cos\theta, \nonumber \\ 
& A_{\theta,33}=\dfrac{\sqrt{3}}{6}\ell \sin\theta +\dfrac{\ell}{2}\cos\theta. \nonumber
\end{align}
Therefore, combining the stability criterion,
Eq.~\eqref{eq:stability_criterion}, with Eqs.~\eqref{eq:total_force2},
\eqref{eq:Atheta}, and \eqref{eq:Atheta-components}, we obtain the
following system of equations
\begin{eqnarray}\label{eq:stability_criterion_calculated}
\left\{
\begin{array}{cc}
\dfrac{2\sqrt{3}}{3} \left(\dfrac{\epsilon}{\ell}\right)F \cos\theta & <f_b \\ \\
-\dfrac{2\sqrt{3}}{3} \left(\dfrac{\epsilon}{\ell}\right)F \cos\left(\theta -\pi/3\right)&< f_b \\ \\
-\dfrac{2\sqrt{3}}{3} \left(\dfrac{\epsilon}{\ell}\right)F \cos\left(\theta +\pi/3\right)&< f_b \\ \\
\end{array},
\right. \\ \nonumber
\end{eqnarray}
where $F=3\pi \epsilon h^{-1} g(\phi) U_\mathrm{av}$ is the total
force exerted by the fluid and the pile on the first layer of grains
($n=1$). The stability criterion,
Eq.~\eqref{eq:stability_criterion_calculated}, can be re-written as
\begin{eqnarray}\label{eq:stability_CEP}
\left\{
\begin{array}{ccc}
1-\dfrac{2\sqrt{3}}{3}\Ca^\star\cos \theta &>0 \\
\\
1+\dfrac{2\sqrt{3}}{3}\Ca^\star\cos\left(\theta -\pi/3\right)&>0 \\
\\
 1+\dfrac{2\sqrt{3}}{3}\Ca^\star\cos(\theta +\pi/3)& >0 \\
\end{array},
\right. \\ \nonumber
\end{eqnarray}
where $\Ca^\star$ is a capillary number that incorporates geometrical
effects of the porous structure and is defined as
\begin{figure}
  \centerline{\includegraphics[width=0.5\textwidth]{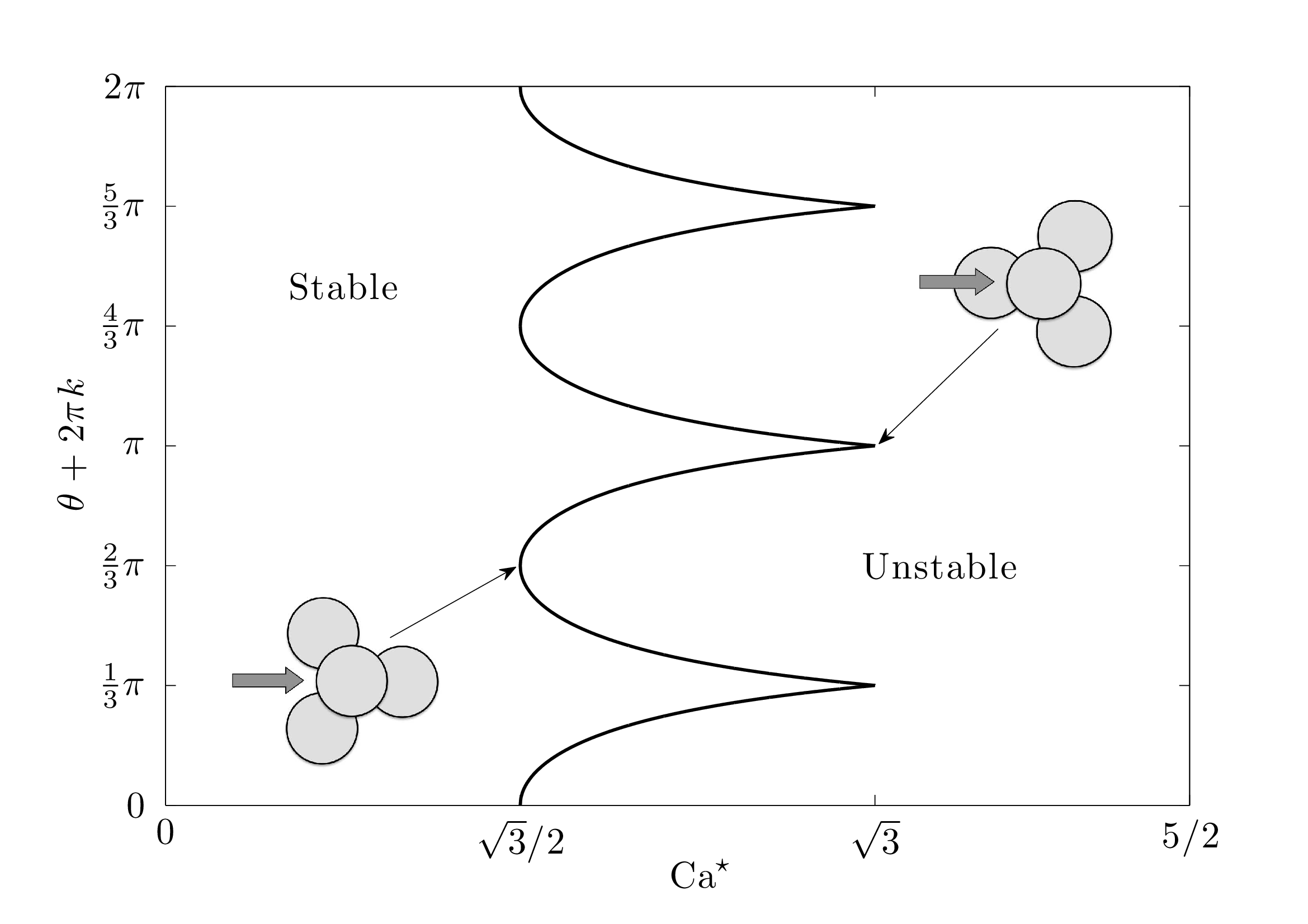}}
  \caption{Fluidization threshold of a wet granulate under
    flow-induced shear in terms of packing orientation $\theta$ and
    capillary number $\Ca^\star$.}
\label{fig:stability-Ca-phi}
\end{figure}
\begin{align}\label{eq:Ca_star}
\Ca^\star=\dfrac{p(\epsilon,\ell) U_\mathrm{av}}{f_b},
\end{align}
where $p(\epsilon,\ell)=3\pi \epsilon^2 g(\phi)/\ell h$ incorporates the impact of pore-scale geometry, and $\phi$ and $h$ are uniquely determined for any fixed $(\epsilon,\ell)$. Assuming a toroidal shape of the liquid surface of the capillary bridges, the dimensional capillary force $\hat f_b=\mu q H f_b$ can be related to the dimensional surface tension $\hat \gamma=\mu q \gamma$ and the contact angle between the wetting liquid (e.g. water) and the surface of the spheres $\eta$ by
\begin{align}\label{eq:capillary-forces}
\hat f_b=2\pi R \hat \gamma \cos \eta
\end{align}
\citep[p. 190]{scheel-2008-morphological}.  Therefore,
\begin{align}
\Ca^\star=\dfrac{p'(\epsilon,\ell)}{ \cos \eta}\Ca,
\end{align}
where $p'=3\epsilon g/(\ell h)$ and $\Ca=\mu \hat U_\mathrm{av}/\hat \gamma$ is the capillary number defined in terms of the average velocity $\hat U_\mathrm{av}$.
Solving Eq.~\eqref{eq:stability_CEP} leads to the following stability criterion for the CEP as a function of packing orientation $\theta$ and capillary number $\Ca^\star$,
\begin{eqnarray}\label{eq:final_stability_CEP}
\left\{
\begin{array}{cccc}
& \Ca^\star<\frac{\sqrt{3}}{2}, & \mbox{stable} &  \forall \theta\\ \\
& \Ca^\star>\sqrt{3},  & \mbox{unstable} & \forall \theta\\ \\
& \Ca^\star \in(\frac{\sqrt{3}}{2},\sqrt{3}), & \mbox{stable if} & \theta \in\left[\mbox{acos}(\sqrt{3}/2\Ca^\star),\right.\\ 
 & & &\,\left.\frac{2}{3}\pi-\mbox{acos}(\sqrt{3}/2\Ca^\star)\right]\\
\end{array}
\right.
\end{eqnarray}

A graphical representation of Eq.~\eqref{eq:final_stability_CEP} is
provided in Figure~\ref{fig:stability-Ca-phi}. The stability of a pile
with CEP arrangement of its grains is affected by the orientation of
the lattice directions relative to the average velocity of the flow by
a factor of 2 (Fig.~\ref{fig:stability-Ca-phi}): the pile orientation
determines how the destabilising hydrodynamic forces decompose along
the lattice directions and, consequently, how they are balanced by the
stabilising capillary forces acting at the contact points.  Combining
Eqs.~\eqref{eq:Ca_star} and~\eqref{eq:capillary-forces}, $\Ca^\star$
can be written as follows
\begin{align}
\Ca^\star=\left(\dfrac{3\mu \epsilon^2 g(\phi)}{2 \hat \gamma\ell h \cos\eta}\right) \dfrac{H}{R}\hat U_\mathrm{av}
\end{align}
Besides the geometrical arrangement  and the physical properties of the wetting and non-wetting fluids (first parenthesis on the RHS), two relevant parameters that control the fluidization threshold are the average velocity across the granulate and the number of layers, since $\Ca^\star$ is proportional to $H/R$. While filtration velocities in low permeability porous media are generally very small, scale effects play a crucial role in determining whether or not flow-induced shear might become a significant source of instability of unconsolidated cohesive granulates.

\section{Summary and Conclusions}
\label{sec:conclusions}

In many environmental and industrial systems, the instability of
cohesive granulates is triggered by a combination of body
(e.g. gravity), surface (e.g. friction) and (boundary and/or
flow-induced) shearing forces. Flow-induced shear forces represent an
important instability factor in many systems where fluid flow occurs,
e.g. cliff instability after heavy rainfall, sediment transport in
submerged environments, and pumping operations during oil recovery,
just to mention a few.  While a number of works have focused on the
effect of friction, gravity and boundary shearing on cohesive
granulates instability, studying the impact of flow-induced shear
forces represents a major challenge since it would a priori require
the full (numerical) solution of Navier-Stokes equations in highly
complex geometry for drag computation.

In this work we use a multi-scale framework to account for the effect
of fluid dynamic shearing on the stability of cohesive granulates.  We
provide closed-form expressions for the instability threshold, due to
flow-induced shear forces, of a regular packing of cohesive
mono-disperse spherical grains in a planar fracture. Without loss of
generality, the analysis is specialised to cohesive forces of
capillary nature. In this setting, the compound effect of structural
(e.g. porosity, grain contacts distribution, pile orientation) and
dynamical (e.g. capillary and fluid-dynamics forces) properties of the
system on its stability is taken into account. The impact of packing
orientation is also quantified: the orientation of a CEP pile affects
its stability threshold by a factor of 2.  Moreover, we identify the
capillary number, $\Ca^\star$, Eq.~\eqref{eq:Ca_star}, as the
dimensionless parameter that controls the instability
threshold. $\Ca^\star$ is the ratio between destabilizing fluid
dynamic shear forces and stabilizing cohesive (capillary) forces,
given by Eqs.~\eqref{eq:total_force} and \eqref{eq:capillary-forces},
respectively. It is defined in terms of the average shearing velocity
$\hat U_\mathrm{av}$, the fluid viscosity $\mu$, the surface tension
$\hat \gamma$, the geometrical arrangement of the grains, the contact
angle $\eta$ between wetting liquid and the surface of the solid
grains, and the ratio between the height of the pile $H$ and the
typical grain diameter $R$. This implies that even though filtration
velocities might be very small, creeping flow might play a key role in
the instability of unconsolidated cohesive granulates due to scale
effects.  We stress that, while applied to cohesive capillary forces,
the method can also be used to model any type of cohesive forces
(e.g. Van der Waals).  Generalisation to include gravity and/or
friction is also straightforward.

Idealized systems as those considered in this study can
provide interesting insights on the salient features, for example,
location of failure, and relevant parameters controlling
cohesive granulates instabilities induced by fluid shear. Also,
since regularly arranged monodisperse granulates lead to a
uniform distribution of the loads, their instability threshold
might provide a sufficient condition for the instability of
similarly loaded/sheared disordered packings where the load
is highly localized to fewer force chains bearing a higher
maximum load.

Real systems, on the other hand, exhibit a host of additional features
including grains polidispersity in size and shape, additional forcing
factors (e.g., gravity, friction, nonuniform distribution of cohesive
forces), and fluid anomalous rheology (i.e., non-Newtonian
behavior). These affect the force balance at both a local and global
level due to structural changes of the force network and contact loads
distribution. In random packings, the latter is highly anisotropic and
inhomogeneous due to the presence of force chains that bear most of
the load. While such inhomogeneity might potentially induce
significant deviations from the behavior of cohesive granulates with a
regular arrangement of grains, it has been shown that
analysis/predictions based upon regular arrangements of frictionless
monodisperse spheres provide remarkably good predictions concerning
the stability properties of nonspherical randomly packed frictional
granulates \cite{Novak-2005-maximum}. This might be attributed to the
observed constant mechanical strength of randomly packed wet
granulates over a wide range of wetting liquid contents
\cite{scheel-2008-morphological}. Scheel et
al. \cite{scheel-2008-morphological} theoretically derived the
cohesivity of randomly packed glass beads by approximating them with
uniform arrangements of frictionless spheres. Their experiments on
both monodisperse and polydisperse sand grains led to a remarkably
good match with their theoretical predictions. It has been therefore
speculated that roughness, as well as randomness, might play only a
secondary role in determining the static and dynamic properties of
random polydisperse granulates \cite{Kudrolli2008}. Fully quantitative investigations
on random (polydisperse) granulates are therefore needed to elucidate
such mechanisms.

The study of contact loads distribution in random networks related to
the onset of instability poses significant additional analytical
challenges since it requires the evaluation of loads spatial
distribution. More importantly, the tails of such distributions have
to be evaluated since they are associated to the maximum loads, which
drive the global instability of the system. While analytical
probability density function (pdf) methods might be employed to obtain
the full pdf of the contact loads, load redistribution to surviving
chains after local rupture could be addressed by, for example, random
fiber bundle model \cite{Dalton2010}. A combination of such analytical and
numerical methods could represent an alternative to computationally
intensive full molecular dynamics simulations.  In addition, many
fluids in natural, industrial, and biological systems exhibit
non-Newtonian behavior (e.g., oil, paints, blood). Anomalous rheology
of the flowing fluid dramatically affects the macroscopic behavior of
the system, its governing equations, and the stress distribution
inside the granulate due to the nonlinear coupling between pore-space
geometry and the rheological properties of the fluid
\cite{Morais2009}. Nevertheless, it has been showed that power-law fluids exhibit
universality behavior, and that their flow properties might belong to
the same universality class of Newtonian fluid flows \cite{Morais2009}. This
suggests that the approach employed in our study could be easily
generalized to non-Newtonian fluids.  The application of pdf methods
to randomly packed cohesive granulates’ instability will be object of
future investigations, together with the study of regular packing
structures other than cubic (e.g., arrangements derived from hexagonal
close packing), size and density polydispersity, and the effect of
fluid non-Newtonicity.

\begin{acknowledgments}
  Funding from BP International within the ExploRe program is
  gratefully acknowledged.
\end{acknowledgments}

\end{document}